\documentclass[conference,10pt]{IEEEtran}
\IEEEoverridecommandlockouts
\usepackage{cite}
\usepackage{amsmath,amssymb,amsfonts}
\usepackage{algorithmic}
\usepackage{graphicx}
\usepackage{textcomp}
\usepackage[table,xcdraw]{xcolor}
\usepackage[a4paper, total={184mm,239mm}]{geometry}
\def\BibTeX{{\rm B\kern-.05em{\sc i\kern-.025em b}\kern-.08em
    T\kern-.1667em\lower.7ex\hbox{E}\kern-.125emX}}

\usepackage[utf8]{inputenc}
\usepackage{braket}
\usepackage{amsmath}
\usepackage{diagbox}
\usepackage{tikz}
\usetikzlibrary{quantikz}

\usepackage{amsthm,amsmath,amssymb}
\usepackage{subfigure}
\usepackage{booktabs}
\usepackage{diagbox}

\usepackage{pgfplots}
\usepackage{multirow}

\theoremstyle{plain}

\newtheorem{prop}{Proposition}
\newtheorem{exam}{Example}

\theoremstyle{remark}

\theoremstyle{definition}
\newtheorem{defn}{Definition}

\usepackage{algorithm}
\usepackage{algorithmic}

\newcommand{\h}{\mathcal{H}}

\newcommand{\blue}[1]{{\color{black}{#1}}}

\usepackage[normalem]{ulem}

\usepackage{todonotes}

\usepackage{authblk}
    
\begin{document}

\title{Image Computation for Quantum Transition Systems
\thanks{Work partially supported by the National Science Foundation of China (12471437), Beijing Nova Program 20220484128, and the Australian Research Council (DP220102059).}
}

\author[1]{Xin Hong}
\author[1]{Dingchao Gao}
\author[2]{Sanjiang Li}
\author[1]{Shenggang Ying}
\author[2]{Mingsheng Ying}
\affil[1]{Key Laboratory of System Software (Chinese Academy of Sciences) and State Key Laboratory of Computer Science}
\affil[ ]{Institute of Software, Chinese Academy of Sciences, Beijing, China}
\affil[2]{Centre for Quantum Software and Information, University of Technology Sydney, Sydney, Australia}
\affil[ ]{Emails: hongxin@ios.ac.cn, gaodc@ios.ac.cn, sanjiang.li@uts.edu.au, yingsg@ios.ac.cn, mingsheng.ying@uts.edu.au}
\renewcommand*{\Affilfont}{\small\it} 
\renewcommand\Authands{ and }
\setlength{\affilsep}{0.5em}

% \author{\IEEEauthorblockN{1\textsuperscript{st} Xin Hong}
% \IEEEauthorblockA{\textit{Key Laboratory of System Software (Chinese Academy of Sciences) and State Key Laboratory of Computer Science} \\
% \textit{Institute of Software, Chinese Academy of Sciences}\\
% Beijing, China \\
% hongxin@ios.ac.cn}
% \and
% \IEEEauthorblockN{2\textsuperscript{nd} Dingchao Gao}
% \IEEEauthorblockA{\textit{Key Laboratory of System Software (Chinese Academy of Sciences) and State Key Laboratory of Computer Science} \\
% \textit{Institute of Software, Chinese Academy of Sciences}\\
% Beijing, China \\
% gaodc@ios.ac.cn
% }
% \and
% \IEEEauthorblockN{3\textsuperscript{rd} Sanjiang Li}
% \IEEEauthorblockA{\textit{Centre for Quantum Software and Information} \\
% \textit{University of Technology Sydney}\\
% Sydney, Australia \\
% sanjiang.li@uts.edu.au}
% \and
% \IEEEauthorblockN{4\textsuperscript{th} Shenggang Ying}
% \IEEEauthorblockA{\textit{Key Laboratory of System Software (Chinese Academy of Sciences) and State Key Laboratory of Computer Science} \\
% \textit{Institute of Software, Chinese Academy of Sciences}\\
% Beijing, China \\
% yingsg@ios.ac.cn
% }
% \and
% \IEEEauthorblockN{5\textsuperscript{th} Mingsheng Ying}
% \IEEEauthorblockA{\textit{\ \ \ \ \ \ \ \ \ \ \ \ \ \ \ \ \ \ \ \ \ \ \ \ \ \ \ \ \ \ \ \ \ \ \ \ \ \ \ \ \ \ \ \ \ \ \ \ Centre for Quantum Software and Information\ \ \ \ \ \ \ \ \ \ \ \ \ \ \ \ \ \ \ \ \ \ \ \ \ \ \ \ \ \ \ \ \ \ \ \ \ \ \ \ \ \ \ \ \ \ \ \ } \\
% \textit{University of Technology Sydney}\\
% Sydney, Australia \\
% mingsheng.ying@uts.edu.au}
% }

\maketitle

\begin{abstract}
With the rapid progress in quantum hardware and software, the need for verification of quantum systems becomes increasingly crucial. While model checking is a dominant and very successful technique for verifying classical systems, its application to quantum systems is still an underdeveloped research area. This paper advances the development of model checking quantum systems by providing efficient image computation algorithms for quantum transition systems, which play a fundamental role in model checking. In our approach, we represent quantum circuits as tensor networks and design algorithms by leveraging the properties of tensor networks and tensor decision diagrams. Our experiments demonstrate that our contraction partition-based algorithm can greatly improve the efficiency of image computation for quantum transition systems.
\end{abstract}


\section{Introduction}
The past several years have seen rapid progress in quantum hardware and software, which makes the need for verification of quantum systems increasingly pressing. Indeed, a series of testing and verification techniques have been developed for quantum circuits in the past 15 years. Several previous works have studied the equivalence checking \cite{Viamontes07iccad-checking,Burgholzer21tcad-advanced,Hong21dac-approx,Burgholzer22dac-handling} and testing \cite{burgholzer2021random} of quantum circuits.

% Moreover, model checking techniques have been developed for verification and analysis of quantum cryptographic protocols \cite{ardeshir2014verification,gay2008qmc}, quantum logic circuits \cite{ying21fm-model}, and quantum programs \cite{Ying21mcqsbook}.

Model checking is an important formal method for verifying finite-state systems. It models the system under examination as a transition system and specifies (safety or liveness) properties in a temporal logic. Given a model $M$ and a property $\varphi$, it then checks if $M$ satisfies $\varphi$. This is usually done by computing all its reachable states through the repeated use of an image computation procedure.

% This is usually done automatically, and a counterexample will be generated when the answer is `no'. After 40 years of development, model checking has become one predominant verification method and has found many successful applications in both hardware and software. 
% The advanced symbolic model checking \cite{McMillan93} represents sets of states and transition relations as reduced ordered binary decision diagrams (ROBDDs or simply BDDs) \cite{bryant1986graph} and checks if a model satisfies a temporal property by computing all its reachable states, which is done by calling an image computation algorithm iteratively until a fixpoint is reached. 

Image computation calculates the output (image) of a set of states under a transition system. Through the repeated calling of this procedure, we can obtain all the reachable states of this system and check the corresponding properties.
Many efficient image computation algorithms have been developed for classical model checking. In particular, image computation can be significantly sped up by taking advantage of the symbolic representation of initial states and transition relations as BDDs \cite{burch1994symbolic} and by exploiting state space partition \cite{Cho_1996} and circuit partition \cite{burch1991symbolic}. Image computation has also been implemented in various classical verification tools (e.g., VIS \cite{brayton1996vis}). 

Model checking techniques have also been extended to quantum systems. Some earlier works focus on verification of quantum communication protocols \cite{gay2005probabilistic}, and some others were motivated by termination analysis of quantum programs \cite{feng2015qpmc}. We refer the reader to the recent book \cite{Ying21mcqsbook} for a systematic introduction to the principle and algorithms for model checking quantum systems. Notwithstanding these achievements, applications of model checking to quantum hardware verification are still an underdeveloped research area. Recently, a temporal extension of the Birkhoff-von Neumann quantum logic \cite{birkhoff1936logic} was proposed in \cite{ying21fm-model}, in which atomic propositions are represented as (closed) subspaces of the quantum state space. While very simple, some important temporal properties of quantum circuits can be specified in this logic and checked by simulation on classical computers \cite{ying21fm-model}. However, one challenge is that we are still lacking efficient implementations of the algorithms for checking quantum circuits. 

This paper aims to advance the development of model checking quantum systems by providing efficient image computation algorithms for quantum systems. Similar to classical model checking, we represent each quantum system as a quantum transition system, where sets of states are replaced by subspaces and transition relations are replaced by quantum operations. In our approach, we regard quantum circuits as tensor networks and design algorithms by leveraging the properties of tensor networks and tensor decision diagrams (TDDs) \cite{hong2020tensor}, which are data structures that, similar to BDDs, represent tensors in a canonical and compact way. More precisely, we first introduce a basic algorithm for conducting image computation for quantum systems and then provide several partition-based optimisation schemes, which play similar roles as the disjunctive and conjunctive partitions for classical image computation \cite{burch1991symbolic}.  
Our experiments demonstrate that our tensor network partition-based algorithms can significantly improve the efficiency of image computation for quantum transition systems when compared with the basic algorithm.

The remainder of this paper proceeds as follows. In section~\ref{sec:pre} we recall relevant background such as quantum states, subspaces, quantum operations, and tensor decision diagrams. The notion of quantum transition systems is then introduced in section~\ref{sec:def}. After describing the basic procedure for conducting the image computation in section~\ref{sec:alg}, we introduce two tensor network partition-based schemes in section~\ref{sec:opt}. 
% Two further optimisation techniques are then introduced in section~\ref{sec:app}. 
Empirical evaluations of these schemes are presented in \ref{sec:exp}. The last section concludes the paper.

\section{Background} \label{sec:pre}

\subsection{Quantum State Space and Quantum Operations}

Quantum states are unit vectors in a Hilbert space $\mathcal{H}$. For a single-qubit system, we write $\h_2$ for its (2-dim) state space. A state in $\h_2$ has form $\ket{\psi}=\alpha \ket{0}+\beta \ket{1}$, where $\alpha$ and $\beta$ are complex numbers and $|\alpha|^2+|\beta|^2=1$. We write $\ket{\pm}$ for the superposition states $\frac{1}{\sqrt{2}}(\ket{0}\pm \ket{1})$. The Hilbert space $\h$ of an $n$-qubit system is the tensor product of $n$ copies of $\mathcal{H}_2$, i.e., $\h=\mathcal{H}_2^{\otimes n}$. A \emph{mixed state} in $\h$ is a density operator $\rho=\sum_i{p_i \ket{\psi_i}\bra{\psi_i}}$ on $\h$, which is a positive semi-definite operator on $\mathcal{H}$ with trace 1. 
% We write $\mathcal{D}(\mathcal{H})$ for the set of all density operators in $\mathcal{H}$. 
The \emph{support} of a density operator $\rho$, denoted as $supp(\rho)$, is the subspace of $\mathcal{H}$ spanned by the eigenvectors corresponding to non-zero eigenvalues of $\rho$.

%A (pure) quantum state is normally represented as a unit vector in a Hilbert space $\mathcal{H}$. A single qubit state, corresponding to the 2-dimensional Hilbert space $\mathcal{H}_2$, can be in the form $\ket{\psi}=\alpha \ket{0}+\beta \ket{1}$, where $\alpha$ and $\beta$ are complex numbers and $|\alpha|^2+|\beta|^2=1$. We will use $\ket{\pm}$ to represent the state $\frac{1}{\sqrt{2}}(\ket{0}\pm \ket{1})$. For an $n$-qubit system, the Hilbert space $\mathcal{H}=\mathcal{H}_2^{\otimes n}$, i.e., the tensor product of $n$ copies of $\mathcal{H}_2$. A mixed state consists of an ensemble of pure states $\{(\ket{\psi_i}, p_i)\}$ and can normally be represented by a density operator $\rho=\sum_i{p_i \ket{\psi_i}\bra{\psi_i}}$, which is a positive operator in $\mathcal{H}$ with trace 1. We use $\mathcal{D}(\mathcal{H})$ to represent the set of all density operators in $\mathcal{H}$. The support of a density operator $supp(\rho)$ is defined as the subspace of $\mathcal{H}$ spanned by the eigenvectors corresponding to non-zero eigenvalues.

Pure quantum states can be transformed by unitary matrices, also called quantum gates, and mixed states are transformed by super-operators. A \emph{super-operator} on $\mathcal{H}$ is a linear operator $\mathcal{E}$ and a \emph{quantum operation} on $\mathcal{H}$ is a completely positive super-operator on $\mathcal{H}$ such that $tr(\mathcal{E}(\rho)) \leq tr(\rho)$ for any density operator $\rho$ in $\mathcal{H}$. Using the Kraus operator-sum representation, a super-operator can be represented as a set of operators $\mathcal{E}=\{E_i\}$ such that $\mathcal{E}(\rho)=\sum_i{E_i\rho E_i^{\dagger}}$.

%A pure quantum state can be transformed by unitary matrices, which are also described by quantum gates, and a mixed state can be transformed by super-operators. A super-operator in $\mathcal{H}$ is a linear operator $\mathcal{E}$ in $\mathcal{L(H)}$ – the space of (bounded) operators in $\mathcal{H}$. A quantum operation in $\mathcal{H}$ is a complete positive super-operator in $\mathcal{H}$ such that $tr(\mathcal{E}(\rho)) \leq tr(\rho)$ for any density operator $\rho$ in $\mathcal{H}$. Using the Kraus operator-sum representation, a super-operator can also be represented as a set of operators $\mathcal{E}=\{E_i\}$ in $\mathcal{H}$ such that $\mathcal{E}(\rho))=\sum_i{E_i\rho E_i^{\dagger}}$. A quantum circuit consists of a set of quantum gates.

% Let us first introduce several notations. For each density operator $\rho$ on $\mathcal{H}$, its support $supp(\rho)$ is defined as the subspace of $\mathcal{H}$ spanned by the eigenvectors corresponding to non-zero eigenvalues.

\begin{defn}
Given a subspace $S$ of $\mathcal{H}$. Its \emph{image} under a set of quantum operations $\mathcal{T}=(\mathcal{T}_\sigma)_{\sigma \in \Sigma}$ is defined as
$\mathcal{T}(S)=\bigvee_{\sigma \in \Sigma}{\mathcal{T}_{\sigma}(S)}$, 
where for each $\sigma \in \Sigma$:
$$\mathcal{T}_\sigma(S)=\bigvee_{\ket{\psi} \in S}{supp(\mathcal{T}_\sigma(\ket{\psi}\bra{\psi}))}.$$
\end{defn}  

Recall that the \emph{join} of a family of subspaces $\{X_i\}$ of $\mathcal{H}$ is
defined as $\bigvee_{i}{X_i}=span(\bigcup_{i}{X_i})$. \blue{Thus, $\mathcal{T}(S)$ is still a subspace.}

The following results are useful in our work.
\begin{prop}[\cite{Ying21mcqsbook}] \label{prop:1}
Let $\mathcal{T}$ be a quantum operation. Then
\begin{enumerate}
    \item $\mathcal{T}(\bigvee_{i}{S_i})=\bigvee_{i}{\mathcal{T}(S_i)}$.
    \item If $\mathcal{T}=(\mathcal{T}_\sigma)_{\sigma \in \Sigma}$ and each $\mathcal{T}_{\sigma}$ has the Kraus operator-sum representation $\mathcal{T}_{\sigma}= \{ E_{\sigma j_\sigma} \}$,
then
$$\mathcal{T}(S)=span\Big(\bigcup_{\sigma,j_\sigma}{\{E_{\sigma j_\sigma}\ket{\psi}:\ket{\psi}\in S\}}\Big).$$
\end{enumerate}
\end{prop}

Combining 1) and 2), $\mathcal{T}(S)$ can also be written as:
$$\mathcal{T}(S)=span\Big(\bigcup_{\sigma,j_\sigma}\left\{E_{\sigma j_\sigma}\ket{\psi}:\ket{\psi}\in B\right\}\Big),$$
where $B=\{\ket{\psi_1},\cdots,\ket{\psi_k}\}$ is a  basis of the subspace $S$.

\subsection{Tensor Decision Diagram}\label{subsec:tdd}
A tensor $\phi_{x_1,\cdots,x_n}: \{0,1\}^I\to\mathbb{C}$ is a multi-dimensional map, where $I=\{x_1,\cdots,x_n\}$ is the set of indices and $n$ is called its rank. Matrices and, thus, quantum operations are examples of tensors. In this work, we assume that every index can take two different values 0 and 1 and will only consider $2^n\times 2^n$ matrices, which are regarded as rank $2n$ tensors.

\renewcommand{\arraystretch}{1.2} % adjust row spacing
\setlength{\arraycolsep}{2pt} % adjust column separation

\begin{figure}[tbh]
\vspace*{-5mm}
  \centering
  \scalebox{0.8}{
  \begin{minipage}{0.3\textwidth}
  \small
\[P=\frac{1}{6}\begin{bmatrix}
1&-1  &1&-1  & 1&-1 &0&0\\
-1&1  &-1&1  & -1&1 &0&0\\
1&-1  &1&-1  & 1&-1 &0&0\\
-1&1  &-1&1  & -1&1 &0&0\\
1&-1  &1&-1  & 1&-1 &0&0\\
-1&1  &-1&1  & -1&1 &0&0\\
0&0   &0&0   &0&0   &3&-3\\
0&0   &0&0   &0&0   &-3&3\\
\end{bmatrix}
\]
  \end{minipage}
 }
 \quad
 \hfil
  \begin{minipage}{0.1\textwidth}  \includegraphics[height = 4cm]{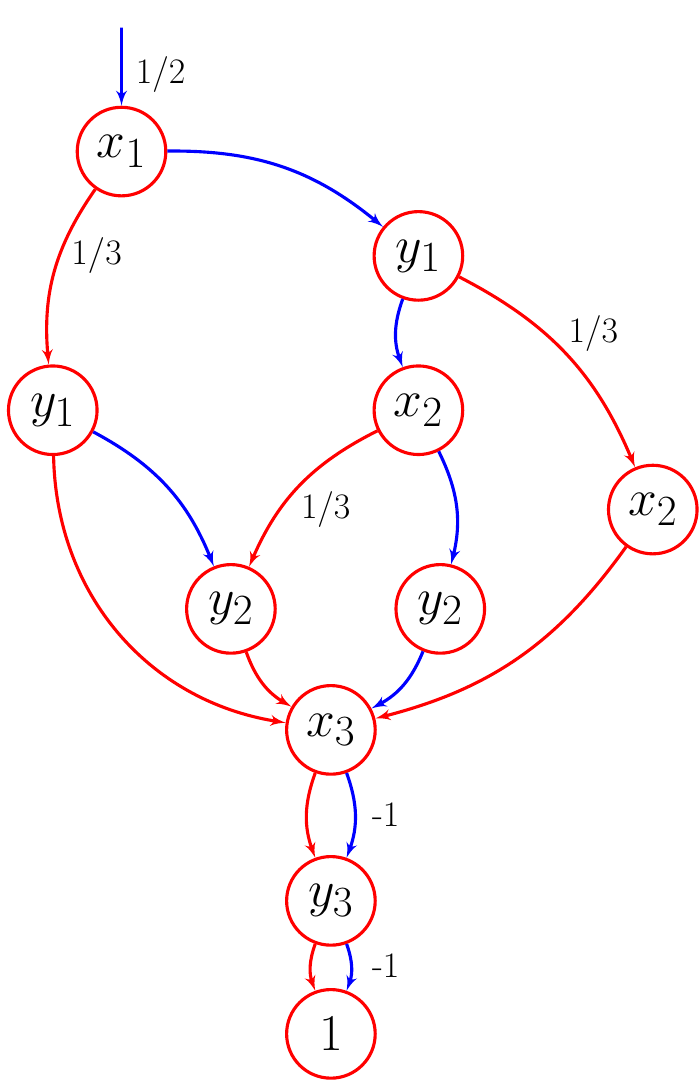}
  \end{minipage}
\caption{A matrix $P$ and its TDD representation. Note that edges with weight 0 are omitted in this diagram.}
    \label{fig:Projector}
\end{figure}

%A tensor is a multi-dimensional array, which can also be seen as a multi-dimensional map from $\{0,1\}^I$ to $\mathbb{C}$. Here, $I=\{x_1,\cdots,x_n\}$ is the set of indices; $n$, the number of indices, is the rank of the tensor, and we sometimes denote the tensor as $\phi_{x_1,\cdots,x_n}$. Matrices and thus quantum gates are common examples of tensors. In our representation, we assume that every index can take two different values 0 and 1. We will see a $2^n\times 2^n$ matrix as a rank $2n$ tensor rather than a rank 2 tensor with every index taking $2^n$ different values. This is achieved by bipartition the matrix according to its column and row number.

Take the matrix $P$ in the left of Fig.~\ref{fig:Projector}
%$$P=\frac{1}{6}\begin{bmatrix}
%1&-1  &1&-1  & 1&-1 &0&0\\
%-1&1  &-1&1  & -1&1 &0&0\\
%1&-1  &1&-1  & 1&-1 &0&0\\
%-1&1  &-1&1  & -1&1 &0&0\\
%1&-1  &1&-1  & 1&-1 &0&0\\
%-1&1  &-1&1  & -1&1 &0&0\\
%0&0   &0&0   &0&0   &3&-3\\
%0&0   &0&0   &0&0   &-3&3\\
%\end{bmatrix}$$
as an example. $P$ can be seen as a rank 6 tensor $\phi_{x_1x_2x_3y_1y_2y_3}$, where the indices $x_i$ and $y_j$ encode the column and row numbers, respectively. For examples, the 7-th column is specified as $(x_1,x_2,x_3)=(1,1,0)$ and the 8-th row is specified as $(y_1,y_2,y_3)=(1,1,1)$. As for the entries, we have $\phi_{x_1x_2x_3y_1y_2y_3}(110111)=-3/6$.

Tensor decision diagrams (TDDs) \cite{hong2020tensor} are designed to represent tensors compactly and canonically. For a fixed index order, each tensor $\phi_{x_1,\cdots,x_n}$ has a unique TDD representation. In a TDD representation, each node (except the terminal node) represents an index, and we use blue and red lines to indicate two different values of the index. Furthermore, each path starting from the root node to the terminal node corresponds to an assignment $a_1,\ldots,a_n$ of the indices $x_1,\ldots,x_n$, and the multiplication of weights along the path corresponds to the tensor value when $x_1=a_1,\ldots,x_n=a_n$. For example, 
in the TDD shown in Fig.~\ref{fig:Projector} (right), the value $\phi_{x_1x_2x_3y_1y_2y_3}(110111)=-1/2=1/2\times 1\times 1\times 1\times 1\times 1\times (-1)$ is obtained by multiplying the weight on the incoming edge of the root node and the weights on the path whose edges are coloured blue, blue, blue, blue, red, blue. 

% Figure \ref{fig:TDD_exp} gives the TDD representation of a Hadamard gate, which can also be seen as a rank two tensor $\phi_{x_1,x_2}=1/\sqrt{2}\cdot (-1)^{x_1\cdot x_2}$. Thus the TDD representation has two levels of internal nodes, labelled with indices $x_1$, and $x_2$, respectively. Each index can take two different values. Thus, every node but the terminal one has two successors, which indicates the positions of the corresponding elements in the tensor. The values of the tensor are then represented by the multiplication of the weights along a path of the TDD. 

%\begin{figure}
%\begin{minipage}[b]{.3\textwidth}
%    \centering
%\small
%\[P=\frac{1}{6}\begin{bmatrix}
%1&-1  &1&-1  & 1&-1 &0&0\\
%-1&1  &-1&1  & -1&1 &0&0\\
%1&-1  &1&-1  & 1&-1 &0&0\\
%-1&1  &-1&1  & -1&1 &0&0\\
%1&-1  &1&-1  & 1&-1 &0&0\\
%-1&1  &-1&1  & -1&1 &0&0\\
%0&0   &0&0   &0&0   &3&-3\\
%0&0   &0&0   &0&0   &-3&3\\
%\end{bmatrix}
%\]
%\end{minipage}
%\begin{minipage}[b]{.45\textwidth}
%    \includegraphics[width=\textwidth]{figures/Projector.pdf}
%\end{minipage}
%\caption{TDD of projector to the space $span\{\ket{++-},\ket{11-}\}$.}
%    \label{fig:Projector}
%\end{figure}

% \begin{figure}
%     \centering
%     \includegraphics[width=0.36\textwidth]{figures/TDD_H_gate2.png}
%     \caption{TDD of the $H$ gate. All the edges are in downward directions implicitly.}
%     \label{fig:TDD_exp}
% \end{figure}

Quantum circuits can be naturally regarded as tensor networks, i.e., sets of tensors connected by shared indices. The most basic operation of a tensor network is the contraction, which sums up two connected tensors over shared indices. Take two tensors $\phi_{xy}$ and $\psi_{yz}$ as an example; their contraction is the rank 2 tensor $\xi_{xz}=\sum_{y=0}^1{\phi_{xy}\cdot \psi_{yz}}$. The contraction as well as the addition of two tensors can be calculated using TDDs directly. Another useful tensor operation is \emph{slicing}: The slicing of a tensor $\phi_{x,y_1,\ldots,y_k}$ with respect to $x=c$ with $c\in\{0,1\}$ is a tensor with index set $\set{y_1,\ldots,y_k}$ such that $\phi|_{x=c}(a_1,\ldots,a_k)=\phi(c,a_1,\ldots,a_k)$.

This paper will use TDD to represent quantum states, quantum operations as well as subspaces. This makes it convenient for leveraging the advantages of both decision diagrams and tensor networks.

\section{Quantum Transition Systems}\label{sec:def}
 A transition system $M$ is a 4-tuple $(S, S_0, \Sigma, R)$, where $S$ is a set of states, $S_0 \subseteq S$ is the set of initial states, $\Sigma=\{\sigma_1,\ldots,\sigma_m\}$ is a set of symbols, and $R \subseteq S \times \Sigma \times S$ is a transition relation. Given a set of states $S' \subseteq S$, its \emph{image} under the transition relation $R$ is defined as 
\begin{equation}\label{eq:image} 
R(S') := \{ t\in S \mid (s, \sigma, t) \in R\ \text{and}\ s \in S'\ \text{and}\ \sigma \in \Sigma\}.
\end{equation}

Quantum transition systems are the quantum generalisation of transition systems and have been introduced in the quantum computing and communication literature with several different names (e.g., quantum automata and (discrete-time) quantum Markov systems). It can be conveniently used to model quantum communication channels and quantum cryptographic protocols \cite{ardeshir2014verification,gay2008qmc}, quantum
circuits \cite{ying21fm-model} and semantics of quantum programs \cite{Ying21mcqsbook}.

%Now we assume the present and next state variables are qubits (i.e. quantum bits) $q = q_1,\cdots, q_n$ and $r = r_1,\cdots, r_n$, respectively, but the input variables are the same $\sigma = \sigma_1, \cdots, \sigma_m$ as in a classical transition system (and thus are classical variables). The state space of every present state variable $q_i$ or next state variable $r_i$ is the same, the 2-dimensional Hilbert space $\mathcal{H}_2$. Then the state spaces of both $q$ and $r$ are $\mathcal{H}=\mathcal{H}_2^{\otimes n}$.

% tensor products $\mathcal{H}=\mathcal{H}_2^{\otimes n}$ of $n$ copies of $\mathcal{H}_2$.

% An $n$-qubit (pure) quantum state $\ket{\psi}$ is a unit vector in $\mathcal{H}$, and an $n$-qubit density operator $\rho$ is a positive operator in $\mathcal{H}$ with trace 1, which normally has the form $\rho= \sum_i{p_i\ket{\psi_i}\bra{\psi_i}}$ corresponding to the ensemble of pure states $\{(\ket{\psi_i},p_i)\}$. A super-operator in $\mathcal{H}$ is a linear operator $\mathcal{E}$ in vector space $\mathcal{L(H)}$ – the space of (bounded) operators in $\mathcal{H}$. A quantum operation in $\mathcal{H}$ is a complete positive super-operator in $\mathcal{H}$ such that $tr(\mathcal{E}(\rho)) \leq tr(\rho)$ for any density operator $\rho$ in $\mathcal{H}$.

% {\color{blue}Ying: Recall the definition of quantum operation (super-operator) here.}

\begin{defn} \label{dfn:qts}
Let $\h$ be a Hilbert space. A quantum transition system $M$ on $\h$ is a 4-tuple $(\mathcal{H}, S_0, \Sigma, \mathcal{T})$, where $S_0$ is a (closed) subspace of $\mathcal{H}$, called the initial space, $\Sigma=\{\sigma_1,\ldots,\sigma_m\}$ is a set of (classical) symbols, and $\mathcal{T}=(\mathcal{T}_\sigma)_{\sigma \in \Sigma}$ is a family of quantum operations on $\mathcal{H}$.
\end{defn}  

For each symbol ${\sigma \in \Sigma}$ a quantum operation $\mathcal{T}_\sigma$ is enabled, which maps a mixed state $\rho$ to another mixed state $\mathcal{T}_\sigma (\rho)$. If the system is closed (i.e., without interaction with its environment), then each $\mathcal{T}_\sigma$ is a unitary transformation $U_\sigma$, which maps a state $\ket{\psi}\in\h$ to another state $U_\sigma \ket{\psi}\in\h$.

% \begin{exam}
% A single-qubit system may suffer from two possible noises: bit-flip and phase-flip. Suppose we cannot decide exactly which noises will occur. The system can be described as a quantum transition system $M=\big(\mathcal{H}_2,S_0,\{1,2\},\{\mathcal{T}_1,\mathcal{T}_2\} \big)$, where $S_0$ is a subspace of $\mathcal{H}$,  $\mathcal{T}_1=\{\sqrt{p}I, \sqrt{1-p}X\}$, and $\mathcal{T}_2=\{\sqrt{p}I, \sqrt{1-p}Z\}$. Note here we use the Kraus operator-sum representation, and $I,X,Z$ are, respectively, the identity matrix, pauli $X$, and pauli $Z$ matrix.
% \end{exam}

Quantum circuits are now the standard model for quantum computing. We now show how the behaviour of quantum circuits can be modelled as quantum transition systems and how the functionality can be checked by calculating their image. Our modelling covers combinational, dynamic, and noisy quantum circuits.

\subsection{Modelling Quantum Circuits}\label{sec:grover}

% Quantum circuits are now the standard model for quantum computing. Essentially, a quantum circuit is a sequence of quantum gates and measurement. We now show how functionality of quantum circuits can be modelled as quantum transition systems. Our modelling covers combinational, dynamic, and noisy quantum circuits.

\subsubsection{Combinational Quantum Circuits} In a combinational quantum circuit, each gate represents a unitary operation. 

%Normally, a unitary transformation can be realised using combinational quantum circuits.

\begin{figure}[h]
\centering

\scalebox{0.67}{
\begin{tikzpicture}
\draw (-4.2,0) node{$x_1^1$};
\draw (-2.55,0) node{$x_1^2$};
\draw (-0.5,0) node{$x_1^3$};
\draw (3.2,0) node{$x_1^4$};
\draw (4.4,0) node{$x_1^5$};

\draw (-4.3,-1.2) node{$x_2^1$};
\draw (-2.55,-1.2) node{$x_2^2$};
\draw (-1.4,-1.2) node{$x_2^3$};
\draw (-0.2,-1.2) node{$x_2^5$};
\draw (0.9,-1.2) node{$x_2^6$};
\draw (2.1,-1.2) node{$x_2^7$};
\draw (3.3,-1.2) node{$x_2^8$};
\draw (4.4,-1.2) node{$x_2^9$};

\draw (-4.5,-2.4) node{$x_3^1$};
\draw (4.4,-2.4) node{$x_3^2$};

\node at (0,0) [anchor=north]{
\begin{quantikz}[column sep=0.28cm,row sep=0.6cm]
&\ctrl{1}&\gate{H}&\qw&\gate{X}&\qw&\qw&\qw     &\ctrl{1}&\qw&\qw&\qw     &\gate{X}&\qw&\gate{H}&\qw \\
&\ctrl{1}&\gate{H}&\qw&\gate{X}&\qw&\gate{H}&\qw&\gate{X}&\qw&\gate{H}&\qw&\gate{X}&\qw&\gate{H}&\qw \\
&\gate{X}&\qw     &\qw     &\qw     &\qw     &\qw     &\qw     &\qw     &\qw&\qw&\qw&\qw&\qw&\qw&\qw
\end{quantikz}};
\end{tikzpicture}
}
\caption{The circuit for Grover iteration. It is also a tensor network with indices $x^j_i$, which denotes the $j$-th index on qubit $i$.}% \magenta{(SL: show the indices in the circuit?)}
\label{fig:cir_grover}
\end{figure}
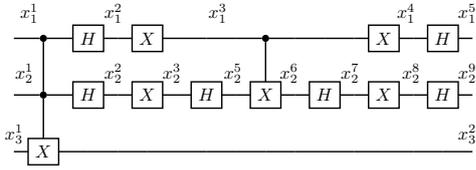

For example, Fig. \ref{fig:cir_grover} gives the circuit for implementing two-qubit Grover iteration \cite{nielsen2002quantum}, a basic procedure of Grover's algorithm for finding a solution of a Boolean function $f(x)=1$. For this circuit, the first CCX gate represents an oracle $O\ket{x}\ket{y}=\ket{x}\ket{f(x)\oplus y}$ and the other gates represent a reflection $2\ket{\psi}\bra{\psi}-I$, where $f(x)=x_1 \wedge x_2$ and $\ket{\psi}=\frac{1}{\sqrt{2}^{n}}\sum_{i=0}^{2^n-1}\ket{i}$. Given an input state $\ket{++-}=\frac{1}{2}\sum_{i=0}^3\ket{i}\ket{-}$, the circuit first changes the state to $\frac{1}{2}\sum_{i=0}^2 \ket{i}\ket{-}-\frac{1}{2}\ket{11}\ket{-}$, and then to $\ket{11}\ket{-}$. In general, let $S=span\{\ket{++-},\ket{11-}\}$. Then, for any input state $\ket{\varphi} \in S$, the output state will always be in $S$.

Then, the system can be modelled by a quantum transition system $(\mathcal{H}_8, S, \{1\}, \mathcal{T})$, where $\mathcal{T}_1 = {(2\ket{\psi}\bra{\psi}-I) O}$, and the property can be checked by calculating $\mathcal{T}_1(S)=S$.

\subsubsection{Dynamic Quantum Circuits}

% \begin{figure}[h]
%     \centering
%     \scalebox{0.53}{
%     \begin{quantikz}[column sep=0.28cm,row sep=0.15cm]%
%                   &\ctrl{3}&\qw     &\ctrl{5}&\qw     &\qw     &\qw     &\qw     &\gate{X}&\qw&\qw&\qw \\
%                   &\qw     &\ctrl{2}&\qw     &\ctrl{3}&\qw     &\qw     &\qw     &\qw&\gate{X}&\qw&\qw \\
%                   &\qw     &\qw     &\qw     &\qw     &\ctrl{2}&\ctrl{3}&\qw     &\qw&\qw&\gate{X}&\qw \\
% \lstick{$\ket{0}$}&\gate{X}&\gate{X}&\qw     &\qw     &\qw     &\qw     &\meter{}&\cwbend{-3}&\cwbend{-2} \\
% \lstick{$\ket{0}$}&\qw     &\qw     &\qw     &\gate{X}&\gate{X}&\qw     &\meter{}&\cw&\cwbend{-1}&\cwbend{-2} \\
% \lstick{$\ket{0}$}&\qw     &\qw     &\gate{X}&\qw     &\qw     &\gate{X}&\meter{}&\cwbend{-2}&\cw&\cwbend{-1}
%     \end{quantikz}
%     }
%     \caption{The circuit for correcting one qubit bit-flip error.} 
%     \label{fig:cir_bit_flip}
% \end{figure}

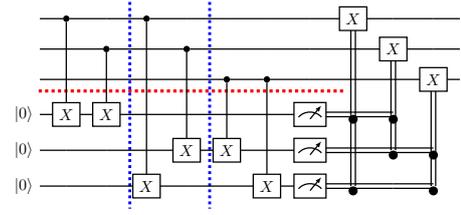
\begin{figure}
    \centering
    \scalebox{0.6}{
    \begin{tikzpicture}
    % \draw[dotted, red, line width =1.5pt] (-4.05,-0.3) rectangle (-2.35,-3.05);
    % \draw[dotted, blue, line width =1.5pt] (-2.24,-0.3) -- (-2.24,-4.6) -- (1.15,-4.6)--(1.15,-1.6)--(0.45,-1.6)--(0.45,-3.9)--(-1.54,-3.9)--(-1.54,-0.3)--(-2.24,-0.3);
    % \draw[dotted, green, line width =1.5pt] (-1.4,-1) rectangle (0.3,-3.8);
    \draw[dotted, red, line width =2pt] (-4.3,-2.1) -- (2.4,-2.1);
    \draw[dotted, blue, line width =1.8pt] (-2.3,0) -- (-2.3,-4.7);
    \draw[dotted, blue, line width =1.8pt] (-0.55,0) -- (-0.55,-4.7);
    \node at (0,0) [anchor=north]{
    \begin{quantikz}[column sep=0.28cm,row sep=0.15cm]%
                  &\ctrl{3}&\qw     &\ctrl{5}&\qw     &\qw     &\qw     &\qw     &\gate{X}&\qw&\qw&\qw \\
                  &\qw     &\ctrl{2}&\qw     &\ctrl{3}&\qw     &\qw     &\qw     &\qw&\gate{X}&\qw&\qw \\
                  &\qw     &\qw     &\qw     &\qw     &\ctrl{2}&\ctrl{3}&\qw     &\qw&\qw&\gate{X}&\qw \\
\lstick{$\ket{0}$}&\gate{X}&\gate{X}&\qw     &\qw     &\qw     &\qw     &\meter{}&\cwbend{-3}&\cwbend{-2} \\
\lstick{$\ket{0}$}&\qw     &\qw     &\qw     &\gate{X}&\gate{X}&\qw     &\meter{}&\cw&\cwbend{-1}&\cwbend{-2} \\
\lstick{$\ket{0}$}&\qw     &\qw     &\gate{X}&\qw     &\qw     &\gate{X}&\meter{}&\cwbend{-2}&\cw&\cwbend{-1}
    \end{quantikz}
    };
    \end{tikzpicture}}
    \caption{The circuit for correcting one qubit bit-flip error.}
    \label{fig:cir_bit_flip}
\end{figure}

%General quantum transitions can be realised by dynamic quantum circuits. 
A dynamic quantum circuit is a quantum circuit in which measurements can happen in the middle and the subsequent circuit can depend on the measurement outcomes \cite{corcoles2021exploiting}. 

Fig. \ref{fig:cir_bit_flip} gives an example of a dynamic quantum circuit which can be used to correct the input quantum state if at most one bit-flip error may occur. For this circuit, the input state will first experience a period of syndrome detection (the six CX gates), denoted by $U$, then a set of measurements will be conducted to identify the error, and lastly, a set of operations will be applied according to the measurement results to correct the error. There are four different measurement results that can happen for this circuit, viz. 000, 101, 110, and 011, corresponding to the cases of no error or a bit-flip error on the first, second, and third qubits, respectively. 

Thus, the system can be modelled by a transition system with four quantum operations $\mathcal{T}_{000}=\{(I_1\otimes I_2\otimes I_3\otimes \ket{000}\bra{000})U\}$, $\mathcal{T}_{101}=\{(X_1\otimes I_2\otimes I_3\otimes \ket{101}\bra{101})U\}$, $\mathcal{T}_{110}=\{(I_1\otimes X_2\otimes I_3\otimes \ket{110}\bra{110})U\}$, $\mathcal{T}_{011}=\{(I_1\otimes I_2\otimes X_3\otimes \ket{011}\bra{011})U\}$, where $I_i$ and $X_i$ are, respectively, the identity operation and $X_i$ on the $i$-th qubit. The correctness of this circuit can be partly checked by $\mathcal{T}(span\{\ket{100},\ket{010},\ket{001}\})=span\{\ket{000}\}$ which means that the bit-flip error has been corrected.

% For this circuit, if the input state has at most one error, e.g., if it is in the subspace $span\{\ket{000},\ket{100}\}$, then the output state will always be  $\ket{000}$, which means that the bit-flip error on the first qubit has been corrected.

\subsubsection{Noisy quantum circuits}
Noises may happen during the execution of quantum circuits. In quantum computing, noises are often represented as super-operators in the Kraus operator sum form.

%Another common example of quantum transition systems is noisy quantum circuits in which noises can happen during the execution of quantum circuits.

% Also, take the Grover iteration as an example; suppose the circuit shown in Fig. \ref{fig:cir_grover} corresponds to the unitary matrix $G$ and a noise bit-flip or phase-flip can happen at the end of the circuit, then the transition system can be represented as $(\mathcal{T}, S)$ where $\mathcal{T}=(\mathcal{G}, \mathcal{T}_1\otimes \mathcal{G}, \mathcal{T}_2\otimes \mathcal{G})$. Here, $\mathcal{G}=\{G\}$ is the super-operator with only one Kraus operator $G$.

\begin{figure}[h]
\centering
\scalebox{0.6}{
\begin{quantikz}[column sep=0.28cm,row sep=0.28cm]
&\gate{H}&\gate[1,style={starburst,draw=yellow,line
width=1pt,inner xsep=-4pt,inner ysep=-5pt}]{\mathcal{N}}&\ctrl{1}&\ctrl{2}&\qw     &\octrl{2}&\octrl{1}&\qw&\qw&\qw \\
&\qw     &\qw               &\gate{X}    &\qw    &\qw     &\qw       &\gate{X}  &\qw&\qw&\qw \\
&\qw     &\qw               &\ctrl{-1}&\gate{X}  &\qw     &\gate{X}   &\ctrl{-1} &\qw&\qw&\qw \\
&\qw     &\qw               &\ctrl{-1}&\ctrl{-1}&\gate{X}&\ctrl{-1} &\ctrl{-1} &\qw&\qw&\qw \\
\end{quantikz}
}
\caption{A noisy version of a quantum walk along a 8-length cycle.}
\label{fig:cir_qrw}
\end{figure}
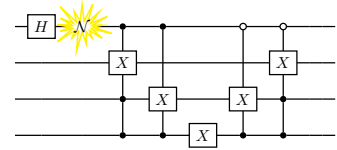

Take a quantum walk along a 8-length cycle \cite{douglas2009efficient, acasiete2020implementation} as an example. Here, the coin operation is implemented by a Hadamard gate 
$H=\frac{1}{\sqrt{2}}
\big[
    \begin{smallmatrix}
        1&1\\
        1&-1
    \end{smallmatrix}
\big]$.
A state will be shifted to an adjacent state and, depending on the value of the coin, it could be either higher or lower. The shift operations can be described by two operators $S_0=\sum_{i=0}^{7}{\ket{(i-1)\!\!\mod 8}\bra{i}}$ and $S_1=\sum_{i=0}^{7}{\ket{(i+1)\!\!\mod 8}\bra{i}}$ and implemented using a series of $C^n(X)$ gates, see Fig.~\ref{fig:cir_qrw}.

\blue{To simplify the discussion, we only consider the case where a noise occurs once here.} Suppose a bit-flip noise may happen after the coin operator. Then the transition system can be represented as $(\mathcal{H}_2^{\otimes 4}, S,\{1,2\},\mathcal{T})$ where $\mathcal{T}_1=\mathcal{S}\circ (\mathcal{E}_c\otimes \mathcal{I})$ and $\mathcal{T}_2=\mathcal{S}\circ (\mathcal{E}_b\otimes \mathcal{I})\circ (\mathcal{E}_c\otimes \mathcal{I}))$. 
% \lsj{What is $\mathcal{T}_1$ here? You also need to explain explicitly on which subsystems $H,I,S$ are acting.}
Here, we use $\mathcal{E}_c$ and $\mathcal{I}$ to represent the super-operator with only one Kraus operator $H$ and $I$, $\mathcal{E}_b=\{\sqrt{p}I, \sqrt{1-p}X\}$ to represent the bit-flip error and $\mathcal{S}$ to represent the super-operator with only one Kraus operator $S_0\oplus S_1$. Both $\mathcal{E}_c$, $\mathcal{E}_b$ are applied on the first qubit; $\mathcal{I}$ is applied on the last three qubits, and $\mathcal{S}$ is applied on all four qubits. In other words, $\mathcal{S}$ applies the operator $S_t$  to the last three-qubit if the state of the control qubit is $\ket{t}$ ($t=0,1$). The properties of this system can be checked by calculating  $\mathcal{T}(span\{\ket{0}\ket{i}\}) = span\{\ket{0}\ket{(i-1)\!\!\mod 8}, \ket{1}\ket{(i+1)\!\!\mod 8}\}$, which means that sometimes a bit-flip error will not influence the reachable subspace significantly. 

% However, if we start from $span\{\ket{+}\ket{i}\}$ the output subspace will changed from $span\{\ket{0}\ket{(i-1)\!\!\mod 8}\}$ to $span\{\ket{0}\ket{(i-1)\!\!\mod 8}, \ket{1}\ket{(i+1)\!\!\mod 8}\}$.

\section{Basic Algorithms}\label{sec:alg}

For classical systems, image computation is normally conducted by representing both the initial set of states and the transition relation as BDDs and then calculating their conjunction and eliminating the existential quantifier \cite{McMillan93, burch1994symbolic}. 

In this section, we describe a basic approach for image computation for quantum systems. We represent Kraus operators $E_i$ TDDs and give algorithms for computing the join of subspaces.

% For a super-operator $\e=\set{E_i}$, we represent each Kraus operator $E_i$ of $\e$ as a TDD. What we need to do next is to represent subspaces as TDDs and design efficient algorithms for computing the join of subspaces.

%In the following part, we discuss the representation of subspaces and other operations.

\subsection{Representation of Subspaces}

% Given a subspace $S$, there are two methods for representing $S$. The first is to represent $S$ as a \emph{projector} $P_S$, or simply $P$, such that a quantum state $\ket{\psi} \in S$ if and only if $P\ket{\psi}=\ket{\psi}$. The second method represents $S$ by a set of basis states $B=\{\ket{\psi_1},\cdots,\ket{\psi_k}\}$, such that any state $\ket{\psi}\in S$ can be represented by the linear combination of $\ket{\psi_1},\cdots,\ket{\psi_k}$. If $B$ forms a standard orthogonal basis, then $P=\sum_{i=1}^k{\ket{\psi_i}\bra{\psi_i}}$.

% For unitary transformations $\set{U_\sigma}$, their images of a subspace $S$ can be calculated as $P_{U_{\sigma}(S)}=U_\sigma P_S U_\sigma^{\dagger}$. However, the join of two subspaces and the image of general transformations have to be calculated using the basis.

\blue{A nonempty Hilbert subspace contains infinitely many possible states.} The most efficient method for representing a subspace is to represent it as a set of basis states. Given the projector $P$ of a subspace $S$, we now show how to find a basis for $S$. Let $P=[\ket{u_1},\cdots,\ket{u_{2^n}}]$ be the column representation of the projector. Then $P\ket{u_i}=\ket{u_i}$ for all $i\in \{1,\cdots 2^n\}$. Thus, if $\ket{u_i} \neq 0$, then $\frac{\ket{u_i}}{\|\ket{u_i}\|}$ can be served as an element in the basis of the subspace, we denote it as $\ket{v_i}$ and update $P=P-\ket{v_i}\bra{v_i}$. Then, we can recursively find a set of orthogonal basis vectors for the original subspace. Although this process may experience a high complexity when traversing all columns using the matrix representation, it can be achieved easily using the TDD representation since the first non-zero column can be found by locating the leftmost non-zero path of the TDD representation of $P$.

\begin{exam}\label{exp:find_basis}
Take the subspace $S=span\{\ket{++-},\ket{11-}\}$ used in Grover iteration (cf. Sec.~\ref{sec:grover})  as an example. The matrix and TDD representations of the corresponding projector $P$ are shown in Fig.~\ref{fig:Projector}. The leftmost path corresponds to $(x_1,x_2,x_3,y_1,y_2,y_3)=(0,0,0,0,0,0)$ while means that the first column of the projector is non-zero. Traverse all paths with $(x_0,x_1,x_2)=(0,0,0)$, we can obtain the vector $\ket{v_1}=\frac{1}{6}[1,-1,1,-1,1,-1,0,0]$, which can be normalised as $\ket{v_1}=\frac{1}{\sqrt{3}}(\ket{00}+\ket{01}+\ket{10})\ket{-}$. Let $P'=P-\ket{v_1}\bra{v_1}$. Then $P'$  equals to $\ket{11-}\bra{11-}$. Repeating the process for $P'$, we obtain $\ket{v_2}=\ket{11}\ket{-}$. Then $\{\ket{v_1},\ket{v_2}\}$ is a basis of the subspace $S$.

% To find a non-zero column vector of $P$, we first identify the leftmost path in the TDD which has no edges with weight 0. Suppose the $x_i$ node on this path has value $a_i$ ($i=1,2,3$). Then we calculate the values for all paths in the TDD whose $x_i$ node has value $a_i$. This gives a nonzero column of $P$. For our example, such a nonzero column is obtained by traversing all paths with $x_0,x_1,x_2$ all being 0. The corresponding vector is $\ket{v_1}=\frac{1}{6}[1,-1,1,-1,1,-1,0,0]$, which can be normalised as $\ket{v_1}=\frac{1}{\sqrt{3}}(\ket{00}+\ket{01}+\ket{10})\ket{-}$. Let $P'=P-\ket{v_1}\bra{v_1}$. Then $P'$  equals to $\ket{11-}\bra{11-}$. Repeating the process for $P'$, we obtain $\ket{v_2}=\ket{11}\ket{-}$. Then $\{\ket{v_1},\ket{v_2}\}$ is a basis of the subspace $S$.
\end{exam}

Note that we do not need to represent the corresponding vectors in the explicit form during the calculation process but to represent them as TDDs, thus reducing the complexity.

% \red{(SL: Here you should explain how to update the TDD.)} \red{(SL: Here you should explain how to update the TDD.)} 
% From the figure, we can see that the first column (corresponding to the paths leading by the 0-successors of $x_0, x_1$ and $x_2$) is nonzero,
% \magenta{(It is not clear how you get this? And how this is related to TDD operations?)} \blue{This is obtained directly from the matrix representation and can also be easily seen from the TDD representation.} 
% \magenta{[Explain more on how this TDD represents the projector.]}

\subsection{Computing the Join of Two Subspaces}

%注，这一步其实很关键，当我们处理含噪声电路时，我们可以把Kraus operator里面的系数p去掉变成酉算子，然后所有的操作都用UPU^{\dagger}去做，但是最后求两个空间的join还是需要先去找到一组基，然后一个一个去加到原始空间，这个是相对复杂的

Let $S=S_1\vee S_2$ be the join of two subspaces $S_,S_2$. Suppose $B_1=\{\ket{\psi_{11}},\cdots,\ket{\psi_{1k}}\}$ and $B_2=\{\ket{\psi_{21}},\cdots,\ket{\psi_{2l}}\}$ are orthonormal bases for $S_1$ and $S_2$. We show how to compute a basis $B$ for $S$. 

Set $B=B_1$ and let $P=\sum_{j=1}^{k}{\ket{\psi_{1j}}\bra{\psi_{1j}}}$. We complete $B$ into a basis of $S$ step by step by following the Gram-Schmidt procedure. That is, we consider basis vectors in $B_2$ one by one. Suppose the current vector is $\ket{\psi_{2j}}$. We calculate $\ket{u_j}=\ket{\psi_{2j}}-P\ket{\psi_{2j}}$ and normalise it as $\ket{v_j}=\frac{\ket{u_j}}{\|\ket{u_j}\|}$. If $\ket{v_j}$ is 0, then we consider the next vector; otherwise, it is orthogonal to $P$ and we add it to $B$. Meanwhile, we also update $P$ as $P+\ket{v_j}\bra{v_j}$. Repeat this process until all elements in $B_2$ have been considered. Then, $B$ will be a basis of $S$ and $P$ will be the projector to $S$. 

%For any $\ket{\psi_{2j}}$, if $\ket{\psi_{2j}} \in S_1$ then it can be deleted from $B_2$; if $\ket{\psi_{2j}} \in S_1^{\perp}$, then we delete it from $B_2$ and add it to $B$. If $B_2$ is still nonempty, then we follow the Gram-Schmidt procedure and generate a basis for $S$.

%Furthermore, we consider obtaining a basis for $S=S_1\vee S_2$ given the basis $B_1=\{\ket{\psi_{11}},\cdots,\ket{\psi_{1k}}\}$, $B_2=\{\ket{\psi_{21}},\cdots,\ket{\psi_{2l}}\}$ for $S_1$ and $S_2$. For any $\ket{\psi_{2j}}$, if $\ket{\psi_{2j}} \in S_1$ then it can be deleted and if $\ket{\psi_{2j}} \in S_1^{\perp}$ it can be added to the set $B_1$ to form the basis of $S$. But there are also elements neither in $S_1$ nor in $S_1^{\perp}$. Thus, we follow the Gram-Schmidt procedure to obtain the basis for $S$.

%Let $P=\sum_{j=1}^{k}{\ket{\psi_{1j}}\bra{\psi_{1j}}}$ be the projector to $S_1$, and let $B=B_1$. Then for any $\ket{\psi_{2j}}$, calculate $\ket{v_j}=\ket{\psi_{2j}}-P\ket{\psi_{2j}}$ and normalise it $\ket{v_j}=\frac{\ket{v_j}}{\|\ket{v_j}\|}$ which is orthogonal to $P$. Then update $P$ to $P=P+\ket{v_j}\bra{v_j}$, and update $B$ to $B=B\cup \{\ket{v_j}\}$ if $\ket{v_j} \neq 0$. Repeat this process until all elements in $B_2$ have been considered. Then, $B$ will be a basis of $S$, and $P$ will be the projector to it. 
% \lsj{One may ask why not using the Gram-Schmidt procedure directly?}

\begin{exam}
Consider Grover iteration again. Let  $P_1,P_2$ be the projectors of two one-dimensional subspaces $S_1,S_2$, which are generated by $B_1=\{\ket{++-}\}$ and $B_2=\{\ket{11-}\}$ respectively. Clearly, $P_1=\ket{++-}\bra{++-}$. We complete $B_1$ into a basis of $S=S_1\vee S_2$. A simple calculation shows that $\ket{u}=\ket{11-}-P_1\ket{11-}=\ket{11-}-\frac{1}{4}\ket{++-}=[-\frac{1}{4},\frac{1}{4},-\frac{1}{4},\frac{1}{4},-\frac{1}{4},\frac{1}{4},\frac{3}{4},-\frac{3}{4}]^T$, which can be normalised as $\ket{v}=-\frac{1}{2\sqrt{3}}(\ket{00}+\ket{01}+\ket{10}-3\ket{11})\ket{-}$. Then $B=\{\ket{++-},\ket{v}\}$ is an orthonormal basis, and $P=P_1+\ket{v}\bra{v}$ is the corresponding projector, which is exactly the one shown in Example~\ref{exp:find_basis}.
\end{exam}

\subsection{A Basic Image Computation Algorithm}

Based on the above algorithms, we now give the basic image computation algorithm for quantum transition systems.

\begin{algorithm} 

	\caption{$Basic\_Image(\h,S,\Sigma,\mathcal{T})$} 
	\label{alg-rs} 
	\begin{algorithmic}[1]
		\REQUIRE A quantum transition system $(\h,S,\Sigma,\mathcal{T})$, where $\mathcal{T}=\{\mathcal{T}_\sigma\mid \sigma\in \Sigma\}$ and  $\mathcal{T}_\sigma=\{E_{\sigma,j_\sigma}\}$
		\ENSURE the projector $P$ of $\mathcal{T}(S)$
		\STATE $P \gets 0$ 
		\STATE $B \gets Basis\_Decompose(S)$
		\STATE $K \gets \cup_{\sigma,j_\sigma}\{E_{\sigma,j_\sigma}\}$
		\FOR{$\ket{\psi}\ in\ B$, $E\ in\ K$}
		\STATE $\ket{\phi} \gets cont(\ket{\psi},E)$
		\STATE $P=P \vee span\{\ket{\phi}\}$
		\ENDFOR
		\RETURN $P$ 
	\end{algorithmic} 

\end{algorithm}
% \magenta{[What is an $e\in E$? Consider use better symbols for $e$ and $E$.]}

The basic procedure is to find a basis of the initial subspace and then calculate $E\ket{\psi}$ for every Kraus operator $E$ and basis state $\ket{\psi}$ (both are represented as TDDs) and then calculate the join of all these states. In this paper, we use $cont(\phi_1,\cdots,\phi_k)$ to represent the contraction of $k$ tensors $\phi_1, \cdots, \phi_k$.

\section{Partition-Based Optimisations}\label{sec:opt}

%In classical image computation, the process is normally optimised by the partitioning of transition relations which usually need huge memory resources to be represented as a single BDD. Two commonly used partitioning methods are the disjunctive partition and the conjunctive partition \cite{burch1991symbolic}, which partition the transition relations to the disjunction and conjunction of many smaller transition relations, respectively. In the following subsections, we consider the partition of quantum circuits.

%Since a quantum circuit can not be simply modelled as a Boolean function, we can not partition it as a disjunction or conjunction of some simpler parts. However, we can partition the circuit into the addition and contraction of many parts. Thus, we call these two methods addition partition and contraction partition, respectively.

%Since representing the transition relation as a monolithic BDD usually needs huge memory resources, in classical model checking, image computation is often optimised by partitioning the transition relation.  Two commonly used methods are the disjunctive and the conjunctive partitions \cite{burch1991symbolic}, which partition the transition relations to the disjunction or conjunction of many smaller transition relations.

Representing a transition relation as a monolithic BDD usually needs huge memory resources. In classical model checking, partition-based  methods are introduced \cite{burch1991symbolic} to optimise image computation by partitioning the transition relation as the disjunction or conjunction of smaller transition relations. Analogously, by leveraging the properties of tensor networks and TDDs, we can partition a quantum circuit (regarded as a tensor network or a TDD) into many parts, which can be added or contracted to recover the functionality of the quantum circuit. The two methods are addressed as, respectively, \emph{addition} and \emph{contraction} partitions. 

%Because a general quantum circuit can not be modelled as a Boolean function, it can not be recovered as the disjunction or conjunction of some simpler parts. However, by leveraging the properties of tensor networks and TDDs, we can partition a quantum circuit (regarded as a tensor network or a TDD) into many parts, which can be added or contracted to obtain the original functionality of the quantum circuit. In this sense,  we also address our methods as \emph{addition} and \emph{contraction} partitions, respectively. 

% \magenta{(SL: but you also introduced DD-based partitions.)}

\subsection{Addition Partition}

% In \cite{burch1991symbolic}, the transition relations are partitioned using two methods. One is to partition a transition relation as a disjunction of several simpler transition relations, called disjunctive partition. Another is to partition a transition relation as a conjunction of several simpler transition relations, called conjunctive partition.

% For quantum systems, since the transition relations are more complex than Boolean functions, we partition them as the addition and contraction of a series of simpler tensors.

Regarding quantum circuits as tensor networks, we first transform (as in \cite{chen2018classical}) a quantum circuit $C$ as an undirected graph $G$ and then partition the circuit according to some vertices with the highest degrees. More specifically, every node in $G$ represents an index of the quantum circuit and two nodes are connected in $G$ if they are the input or output indices of the same gate. In this work, we regard two indices as the same if (i) they are the input and output indices of a diagonal quantum gate; or (ii) they are the input and output indices of a control qubit of a controlled gate. That is, hyper-edges \cite{pednault2020paretoefficient} can appear in $G$. This graph captures the connectivity of the quantum circuit and facilitates the partition of the circuit by slicing some indices with the highest degrees. We call this method \emph{addition partition}.

%For addition partition, we mainly use the method of \cite{chen2018classical}. For \cite{chen2018classical}, a quantum circuit is transformed into an undirected graph which reveals the connectivity of the circuit. Then we will partition the circuit according to the vertexes which have high degrees. More specifically, let $G=(V, E)$ be a graph such that every node of it represents an index of the tensor network corresponding to the quantum circuit. Then two nodes are connected if they are the input or output indices of the same gate. Here, hyper-edges will be used. In other words, the input and output of every qubit of a diagonal quantum gate will share the same indices, and the input and output of every control qubit of a controlled gate will also share the same indices. This graph captures the connectivity of the quantum circuit, and the circuit can be partitioned by slicing the indices with the highest degrees. 

% Then, two indices of the same qubit are connected if they are the input and output indices of a non-diagonal quantum gate; two indices of different qubits are connected if they are the input and output indices of a multi-qubit gate. Then this graph captures the connectivity of the quantum circuit. Note that here, the input and output of every qubit of a diagonal quantum gate will share the same indices. Then the circuit can be partitioned by giving values of the high-degree indices. 

\begin{figure}
    \centering
    \scalebox{0.65}{
    \begin{tikzpicture}
\node[circle, minimum width =10pt , minimum height =10pt ,draw= red] (1) at(0,2){\begin{footnotesize}$x_1^1$\end{footnotesize}};
\node[circle, minimum width =10pt , minimum height =10pt ,draw=black] (2) at(1.7,2){\begin{footnotesize}$x_1^2$\end{footnotesize}};
\node[circle, minimum width =10pt , minimum height =10pt ,draw=blue] (3) at(3.4,2){\begin{footnotesize}$x_1^3$\end{footnotesize}};
\node[circle, minimum width =10pt , minimum height =10pt ,draw=black] (4) at(5.1,2){\begin{footnotesize}$x_1^4$\end{footnotesize}};
\node[circle, minimum width =10pt , minimum height =10pt ,draw=black] (5) at(6.8,2){\begin{footnotesize}$x_1^5$\end{footnotesize}};
\node[circle, minimum width =10pt , minimum height =10pt ,draw=black] (6) at(-0.7,0.6){\begin{footnotesize}$x_3^1$\end{footnotesize}};
\node[circle, minimum width =10pt , minimum height =10pt ,draw=black] (7) at(0.7,0.6){\begin{footnotesize}$x_3^2$\end{footnotesize}};
\node[circle, minimum width =10pt , minimum height =10pt ,draw=blue] (8) at(0,-0.8){\begin{footnotesize}$x_2^1$\end{footnotesize}};
\node[circle, minimum width =10pt , minimum height =10pt ,draw=black] (9) at(1,-0.8){\begin{footnotesize}$x_2^2$\end{footnotesize}};
\node[circle, minimum width =10pt , minimum height =10pt ,draw=black] (10) at(2,-0.8){\begin{footnotesize}$x_2^3$\end{footnotesize}};
\node[circle, minimum width =10pt , minimum height =10pt ,draw=black] (11) at(3,-0.8){\begin{footnotesize}$x_2^4$\end{footnotesize}};
\node[circle, minimum width =10pt , minimum height =10pt ,draw=black] (12) at(4,-0.8){\begin{footnotesize}$x_2^5$\end{footnotesize}};
\node[circle, minimum width =10pt , minimum height =10pt ,draw=black] (13) at(5,-0.8){\begin{footnotesize}$x_2^6$\end{footnotesize}};
\node[circle, minimum width =10pt , minimum height =10pt ,draw=black] (14) at(6,-0.8){\begin{footnotesize}$x_2^7$\end{footnotesize}};
\node[circle, minimum width =10pt , minimum height =10pt ,draw=black] (15) at(7,-0.8){\begin{footnotesize}$x_2^8$\end{footnotesize}};
\draw[dashed] (1) --(2);
\draw[-] (2)--(3) --(4)--(5);
\draw[-] (6)--(7);
\draw[-] (8)--(9)--(10)--(11)--(12)--(13)--(14)--(15);
\draw[dashed] (1)--(6)  (1)--(7);
\draw[-] (8)--(6)  (8)--(7);
\draw[-] (3)--(11)  (3)--(12);
\draw[dashed] (1)--(8);

\end{tikzpicture}
}
    \caption{The undirected graph for Grover iteration (cf. Fig.~\ref{fig:cir_grover}), where $x_i^j$ represents the $j$-th index of the $i$-th qubit of the circuit. 
    % Note that the input and output indices of the control qubits of the $CX$ and Toffoli gate are regarded as the same. Thus, a $CX$ (Toffoli) gate is described by a complete graph with 3 (4) nodes.
    }
    \label{fig:udgraph}
\end{figure}
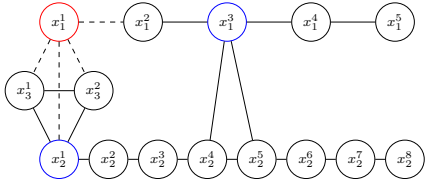

Consider the circuit for Grover iteration (see Fig.~\ref{fig:cir_grover}) again. The corresponding undirected graph is shown in Fig.~\ref{fig:udgraph}. In this circuit, we examine each qubit wire from left to right one by one and use $x_i^j$ to represent the $j$-th index of the $i$-th qubit. From Fig.~\ref{fig:udgraph}, we can see that the nodes labelled with $x_1^1$, $x_2^1$, and $x_1^3$ have the highest degree. We choose any one of the indices and slice the circuit into two simpler ones by giving a value of 0 or 1 to the index.  We can also choose two or all of the indices to partition the circuit into four or eight simpler parts.

% \magenta{[Is $x^i_j$ better?]}
% \magenta{[It is unclear how you partition the circuit.]}

% The circuit for quantum walk (cf. Fig.~\ref{fig:cir_qrw}) is perhaps a more suitable example for demonstrating the advantage of addition partition. Indeed, partitioning the circuit by slicing the coin qubit, we will obtain two much simpler circuits. The undirected graph of this circuit is straightforward to calculate but omitted here due to space restriction.

Suppose a circuit with tensor representation $\phi$ has been partitioned into several parts as above. Let $\phi_1, \cdots, \phi_k$ be their tensor representations. To calculate the image of a state $\ket{\psi}$ under the tensor $\phi$, we need only calculate the contractions $cont(\ket{\psi},\phi_i)$ and then add them up. This is because $cont(\ket{\psi},\phi)=\sum_i{cont(\ket{\psi},\phi_i)}$. 

%The advantage of this method is that we can avoid the appearance of the largest TDD using the basic method. On the other hand, the contraction of different parts $cont(\ket{\psi},\phi_i)$ can be done in parallel.

The advantage of addition partition lies in that calculations of large TDD, as in the basic method, can be avoided. Moreover, contractions of different parts $cont(\ket{\psi},\phi_i)$ can be done in parallel.

\subsection{Contraction Partition}

In contraction partition, we cut the quantum circuit into several smaller parts whose contraction equals the original one. For two preset integer parameters $k_1$ and $k_2$, we partition the circuit into small parts such that every part involves at most $k_1$ qubits and connects with at most $k_2$ multi-qubit gates that across different parts (cf. \cite{hong2020tensor}).%, \magenta{where a cut edge means xxx}

Take the bit-flip code circuit (cf. Fig.~\ref{fig:cir_bit_flip}) as an example. Setting $k_1=3$ and $k_2=2$, the circuit is cut into six blocks as shown in Fig. \ref{fig:cir_bit_flip}. The method first cuts the circuit \emph{horizontally} into $\lceil n/k_1 \rceil$ parts and then cuts the circuit \emph{vertically} whenever $k_2$ multi-qubit gates have been cut, where $n$ is the total number of qubits of the circuit. For this example, $n=6$ and the circuit is cut horizontally into two parts. Whenever the horizontal line cuts two $CX$ gates, we add a vertical line. As a consequence, the circuit is cut into six parts.

Suppose $\phi$ and $\phi_1,\cdots,\phi_k$ are the tensor representations of the circuit and its $k$ blocks. Then $\phi=cont(\phi_1,\cdots,\phi_k)$. To compute the image of a state $\ket{\psi}$ under the transition specified by $\phi$, we need only contract the tensor network connected by these $\phi_i$ and $\ket{\psi}$. By using this method, we can avoid calculating the TDD that represents the whole circuit directly. In most cases, it takes much less memory space to store these smaller TDDs and it takes less time contracting these $\phi_i$ with $\ket{\psi}$ as they have small ranks.

% The two partition schemes can also be used together in the image computation of quantum transition systems.

% Take Grover iteration shown in Fig. \ref{fig:cir_grover} as an example. Given the initial state space $S=span\{\ket{++-},\ket{11-}\}$, we calculate its image under the circuit. As shown in Example~\ref{exp:find_basis}, we first find a basis $\{\ket{v_1},\ket{v_2}\}$ for $S$, where $\ket{v_1}=\frac{1}{\sqrt{3}}(\ket{00}+\ket{01}+\ket{10})\ket{-}$ and $\ket{v_2}=\ket{11}\ket{-}$. We activate the addition partition by slicing $x_1^1$. Applying the first CCX gate, $\ket{v_1}$ will be changed to $\frac{1}{\sqrt{3}}(\ket{00}+\ket{01})\ket{-}$ and $\frac{1}{\sqrt{3}}\ket{10}\ket{-}$ and {$\ket{v_2}$ will be changed to 0}\footnote{This is because the input state $\ket{1}$ vanishes when $x_1^1=0$.} and $-\ket{11}\ket{-}$ respectively in the two slices. We further partition the circuits into two horizontal parts, which involve the first two qubits $[q_0, q_1]$ and the last qubit $[q_2]$, respectively, and then calculate the images of both parts and combine them. The images of $\ket{v_1}$ and $\ket{v_2}$ are $\frac{\sqrt{3}}{2}\ket{v_2}+\frac{1}{2}\ket{v_1}$ and $\frac{1}{2}\ket{v_2}-\frac{\sqrt{3}}{2}\ket{v_1}$, respectively. Thus, the final image is exactly $span\{\ket{v_1},\ket{v_2}\}=S$, which completes the verification of the circuit.

\section{Empirical Evaluation}\label{sec:exp}

% Scalability, 三种方法的对比

% \lsj{Why there are no experiments evaluating the DD-based partitions and the approximation method?}

%In this section, we present experimental results to demonstrate the correctness and efficiency of our algorithms. The correctness can be verified by calculating the image of some circuits with small scales and then comparing them with the theoretical results. In the following part, we focus on efficiency.  All experiments are conducted on a server with an Intel Xeon-Gold-5215 CPU and 512GB RAM.  

%! How can the correctness of an algorithm be verified as above?

In this section, we evaluate the effectiveness and efficiency of our algorithms. All experiments were conducted on a server with an Intel Xeon-Gold-5215 CPU and 512GB RAM. 
% Our algorithms (implemented in Python 3) and benchmarks will be publicly available on the authors' GitHub repository.

%\subsection{Scalability With the Number of Qubits}
\subsection{Comparison Among Three Methods}
% \magenta{(Which one is our preferred method?)}\blue{From the experiment results, the contraction partition algorithm largely outperforms the other two.}
We compare and demonstrate the scalability of our methods on a set of well-known benchmarks. The transition relations of our experiments cover a range of quantum circuits, including circuits for preparing GHZ states and circuits for algorithms such as Grover's algorithm, Bernstein-Vazirani algorithm (BV), quantum Fourier transform (QFT), and quantum random walk (QRW). For the quantum random walk, we add a bit-flip error at the coin qubit after the Hadamard gate. We use the commonly used input states to form the initial subspace of these algorithms. 
% In other words, we set $S_0=span\{\ket{+\cdots+-},\ket{1\cdots1-}\}$ for Grover's algorithm, $S_0=span\{\ket{+\cdots+}\}$ for quantum Fourier transform, and $S_0=span\{\ket{0\cdots0}\}$ for the other three algorithms.

% \lsj{Include more detailed information about these circuits; for example, their input states.}

% the bit flip code circuit, repeat-until-success circuits,

Table~\ref{table:data} presents the experimental results for three image computation methods: the basic algorithm, addition partition, and contraction partition. The first column gives the names of the circuits, which also indicate the qubit number. 
Each of the rest columns corresponds to one of these methods, displaying the time (in seconds) and the maximum node number of TDDs generated during the image computation of the corresponding circuit. For addition partition, we assign a parameter value of $k=1$, which corresponds to dividing the circuit into two parts. 
% This choice is based on the observation that the effectiveness of the addition partition diminishes as the degrees of the nodes being cut decrease. 
For the contraction partition, we set $k_1=k_2=4$, indicating a horizontal circuit cut every four qubits and a vertical cut whenever four multi-qubit gates which have been horizontally cut are encountered.

From Table~\ref{table:data} we can see that all three methods can calculate the images of the 500-qubit BV in 5 minutes and 500-qubit GHZ in 4 seconds. For Grover's algorithm, QFT and QRW, the basic algorithm and addition partition are difficult to go beyond 20 qubits. In general, 
addition partition is better than the basic algorithm. For example, for `QRW\_20', the basic algorithm takes about 341 seconds to finish while addition partition only takes 218 seconds. The maximal node numbers show the same trend. Compared with the basic algorithm and addition partition, our contraction partition method is way more efficient. For example, it takes only 14 seconds for `QRW\_20'. Moreover, it can go far beyond 20 qubits for Grover, QFT, and QRW circuits. More importantly, the maximum TDD size of contraction partition increases at most linearly for QFT, BV, GHZ, and QRW. In the table, we use `-' to indicate that the time exceeds the timeout of 3600s. \blue{It should be noted that, while our method---particularly the contraction partition---has demonstrated good performance in the examples above, the algorithm's overall complexity remains exponential in the worst case.}

\begin{table}[]
\centering
\caption{Experiment data}
\label{table:data}
\scalebox{0.6}{
\begin{tabular}{llllllllll}
\hline
\multirow{2}{*}{Benchmark} &  & \multicolumn{2}{c}{basic} &  & \multicolumn{2}{c}{addition} &  & \multicolumn{2}{c}{contraction} \\ \cline{3-4} \cline{6-7} \cline{9-10} 
                           &  & time        & max \#node       &  & time          & max \#node        &  & time           & max \#node          \\ \hline
% Grover\_15  &  & 23.01   & 16863     &  & 14.96     & 13679 &  & 0.87           & 417      \\
% Grover\_18  &  & 180.89  & 179497     &  & 154.32   &168254 &  & 1.54           & 593   \\
% Grover\_20  &  & 651.27  & 696063     &  & 601.11   &673240 &  & 3.10           & 1178   \\
% Grover\_30  &  &  -      &      &  &  -        &     &  &  61.99         & 36879 \\
% Grover\_40  &  &  -      &      &  &  -        &     &  & 1510.42        & 589865 \\%1405.53
Grover\_15 &   & 19.33  & 15785     &   & 17.35      & 15099  & & 1.61 & 597  \\
Grover\_18 &   & 76.47  & 61694     &   & 66.02      & 60332  & & 2.41 & 516  \\
Grover\_20 &   & 294.65 & 243946    &   & 259.87     & 241240 & & 4.39  & 1036 \\ 
% Grover\_30 &   & -      &           &   & -          &        & & 96.02 & 32785 \\
Grover\_40 &   & -      &           &   & -          &        & & 2953.57 & 851973 \\
\hline
% QFT\_15     &  & 30.87   & 65535     &  & 15.86     & 32770 &  & 0.07           & 61 \\
% QFT\_18     &  & 276.34  & 524287    &  & 141.00    &262146 &  & 0.10           & 61 \\
% QFT\_20     &  & 1264.58 & 2097151   &  & 646.95    &1048578     &  & 0.15           & 64 \\ 
% QFT\_30     &  & -       &      &  & -         &     &  & 0.33           &  61 \\ 
% QFT\_50     &  & -       &      &  & -         &     &  &1.12           & 61    \\ 
% QFT\_100    &  & -       &      &  & -         &     &  &7.74           & 100   \\ 
QFT\_15     &  & 34.64   & 65536   &  & 18.88  & 32770   &  & 0.08 & 63  \\
QFT\_18     &  & 282.12  & 524288  &  & 148.13 & 262146 &   & 0.10  & 31  \\
QFT\_20     &  & 1199.21 & 2097152 &  & 655.19 & 1048578 &  & 0.12 & 63  \\
QFT\_30     &  & -       &         &  & -      &        &  & 0.29 & 31  \\
QFT\_50     &  & -       &         &  & -      &        &  & 1.02 & 51  \\
QFT\_100    &  & -       &         &  & -      &        &  & 7.14 & 101 \\
\hline
BV\_100     &  & 7.36    & 596     &  & 7.43      & 596     &  & 0.41           & 102 \\
BV\_200     &  & 31.57   & 1196    &  & 30.03     & 1196    &  & 1.70           & 202 \\
BV\_300     &  & 75.66   & 1796    &  & 75.56     & 1796    &  & 4.28           & 302 \\
BV\_400     &  & 146.47  & 2396    &  & 145.40    & 2396    &  & 9.18           & 402 \\
BV\_500     &  & 244.15  & 2996    &  & 223.90    & 2996    &  & 16.31          & 502 \\
\hline
GHZ\_100    &  & 0.38    & 595     &  & 0.13      & 301    &  & 0.18           & 200 \\%& 0.03    
GHZ\_200    &  & 0.72    & 1195    &  & 0.37      & 601    &  & 0.48           & 400 \\%& 0.12     
GHZ\_300    &  & 1.29    & 1795    &  & 0.62      & 901    &  & 0.80           & 600 \\%& 0.24     
GHZ\_400    &  & 2.03    & 2395    &  & 1.00      & 1201    &  & 1.26           & 800 \\%& 0.42     
GHZ\_500    &  & 2.96    & 2995    &  & 1.45      & 1501    &  & 1.72           & 1000\\%& 0.62     
\hline
% QRW\_15     &  & 44.14   & 13734     &  & 24.01     & 10878    &  & 3.50           & 222 \\
% QRW\_18     &  & 159.17  & 95074     &  & 84.15     & 37494    &  & 5.45           & 226 \\
% QRW\_20     &  & 457.69  & 278774    &  & 202.47    & 109014    &  & 7.03           & 404 \\
QRW\_15     &  & 36.86   & 13122     &  & 24.59     & 10882     & & 7.16  & 222 \\
QRW\_18     &  & 139.76  & 90538     &  & 84.69     & 37064     & & 11.23 & 226 \\
QRW\_20     &  & 341.05  & 265614    &  & 218.29    & 107714    & & 14.31 & 404 \\
% QRW\_30     &  &  -      &      &  &  -        &     &  & 18.21          &  404\\
% QRW\_50     &  &  -      &      &  &  -        &     &  & 60.38          & 404\\
% QRW\_100    &  &  -      &      &  &  -        &     &  & 351.01         & 428\\
QRW\_30     &   &-       &          &  &-          &          & & 36.82 & 404 \\
QRW\_50     &   &-       &          &  &-          &          & & 118.08 & 404 \\
QRW\_100    &   &-       &          &  &-          &          & & 692.08 & 436 \\
\hline
\end{tabular}}
\end{table}

\subsection{Impact of Parameters for Contraction Partition}

In this subsection, we investigate the influence of the parameters $k_1$ and $k_2$ on the performance of contraction partition. We use the circuit `Grover\_15' as an example and calculate images for different $k_1$ and $k_2$ ranging from 1 to 15. The experimental results are shown in Table~\ref{table:data3}, where the value at entry $(k_1,k_2)$ represents the image computation time for the corresponding parameters $k_1,k_2$ in contraction partition. %\blue{In this table, we have marked times less than 2 seconds in purple and times greater than 5 seconds in blue, with the intensity of the color increasing as the duration becomes longer.} 
\blue{In this table, times under 2 seconds are highlighted in purple, while times exceeding 5 seconds are marked in blue, with the colour intensity increasing as the duration lengthens.} From Table~\ref{table:data3} we can see that the contraction partition method is efficient as long as we do not set the parameters $k_1$, $k_2$ too large. This means that there is a wide range of values that can be chosen for the parameters.

% \begin{figure}[h]
%     \centering

%    \includegraphics[width=0.45\textwidth]{figures/contraction_grover_15_multi.pdf}
 
%     \caption{Experiment data for contraction partition.}

%     \label{fig:data3}
% \end{figure}

% \usepackage{diagbox}
\begin{table}[]
\caption{Time consumption (seconds) for contraction partition}%calculating the image of 'Grover\_15' using contraction partition with different $k_1$ and $k_2$.
\label{table:data3}
\scalebox{0.5}{
	\begin{tabular}{c|ccccccccccccccc}
		\rowcolor[HTML]{FFFFFF} 
		\diagbox{k1}{k2}                         & 1                           & 2                           & 3                           & 4                           & 5                           & 6                           & 7                          & 8                           & 9                           & 10                          & 11                          & 12                          & 13                          & 14                          & 15                          \\\hline
1  & \cellcolor[HTML]{CCC0DA}1.95 & \cellcolor[HTML]{CCC0DA}1.58 & \cellcolor[HTML]{CCC0DA}1.80 & \cellcolor[HTML]{CCC0DA}1.67 & \cellcolor[HTML]{CCC0DA}1.62  & \cellcolor[HTML]{CCC0DA}1.89  & 2.57                          & 2.54                         & 3.64                          & 3.82                          & \cellcolor[HTML]{8DB4E2}6.02  & 3.07                           & 3.28                          & 4.88                          & \cellcolor[HTML]{8DB4E2}6.88  \\
2  & \cellcolor[HTML]{B1A0C7}1.48 & \cellcolor[HTML]{CCC0DA}1.50 & \cellcolor[HTML]{CCC0DA}1.74 & \cellcolor[HTML]{CCC0DA}1.52 & \cellcolor[HTML]{CCC0DA}1.58  & \cellcolor[HTML]{CCC0DA}1.90  & 2.08                          & 2.18                         & 3.01                          & 2.25                          & 4.89                          & 2.68                           & 2.57                          & 4.12                          & \cellcolor[HTML]{8DB4E2}5.63  \\
3  & \cellcolor[HTML]{CCC0DA}1.51 & \cellcolor[HTML]{B1A0C7}1.49 & \cellcolor[HTML]{CCC0DA}1.63 & \cellcolor[HTML]{CCC0DA}1.54 & \cellcolor[HTML]{B1A0C7}1.48  & 2.04                          & 2.29                          & 2.05                       & 3.04                          & 2.17                          & 3.79                          & 2.38                           & 2.59                          & 3.57                          & 3.96                          \\
4  & \cellcolor[HTML]{B1A0C7}1.42 & \cellcolor[HTML]{CCC0DA}1.50 & \cellcolor[HTML]{CCC0DA}1.63 & \cellcolor[HTML]{CCC0DA}1.61                         & \cellcolor[HTML]{CCC0DA}1.62  & 2.20                          & 2.44                          & 2.29                         & 3.04                          & 2.39                          & 3.85                          & 2.91                           & 2.93                          & 3.87                          & 4.86                          \\
5  & \cellcolor[HTML]{CCC0DA}1.65 & \cellcolor[HTML]{CCC0DA}1.65 & \cellcolor[HTML]{CCC0DA}1.75 & \cellcolor[HTML]{CCC0DA}1.64 & \cellcolor[HTML]{CCC0DA}1.90  & 2.14                          & 2.33                          & 2.34                         & 2.86                          & 2.20                          & 3.83                          & 2.58                           & 2.55                          & 3.79                          & 4.29                          \\
6  & \cellcolor[HTML]{CCC0DA}1.69 & \cellcolor[HTML]{CCC0DA}1.74 & \cellcolor[HTML]{CCC0DA}1.71 & \cellcolor[HTML]{CCC0DA}1.75 & \cellcolor[HTML]{CCC0DA}1.60  & 2.02                          & 2.27                          & \cellcolor[HTML]{CCC0DA}1.91 & 2.78                          & 2.14                          & 3.27                          & 2.38                           & 2.93                          & 3.76                          & 3.76                          \\
7  & \cellcolor[HTML]{B1A0C7}1.42 & \cellcolor[HTML]{B1A0C7}1.31 & \cellcolor[HTML]{CCC0DA}1.74 & \cellcolor[HTML]{CCC0DA}1.78 & \cellcolor[HTML]{CCC0DA}1.70  & 3.10                          & 3.41                          & 2.48                         & 4.21                          & 2.36                          & 3.59                          & 3.35                           & 3.43                          & \cellcolor[HTML]{8DB4E2}5.91  & 3.98                          \\
8  & \cellcolor[HTML]{B1A0C7}1.37 & \cellcolor[HTML]{CCC0DA}1.51 & \cellcolor[HTML]{CCC0DA}1.87 & 2.60                         & 2.15                          & 3.45                          & 3.84                          & 3.03                         & 4.60                          & 3.33                          & 4.56                          & 3.70                           & 4.14                          & \cellcolor[HTML]{8DB4E2}6.43  & \cellcolor[HTML]{8DB4E2}5.61  \\
9  & \cellcolor[HTML]{B1A0C7}1.31 & \cellcolor[HTML]{B1A0C7}1.43 & \cellcolor[HTML]{B1A0C7}1.48 & 2.91                         & 2.78                          & 2.83                          & 3.53                          & 4.33                         & 2.86                          & 4.48                          & 3.54                          & 3.12                           & 2.91                          & 4.11                          & \cellcolor[HTML]{8DB4E2}6.03  \\
10 & \cellcolor[HTML]{B1A0C7}1.35 & \cellcolor[HTML]{B1A0C7}1.40 & 2.04                         & 2.60                         & 3.63                          & 3.24                          & 3.19                          & 2.65                         & \cellcolor[HTML]{8DB4E2}5.14  & 3.37                          & 3.21                          & 3.49                           & 4.14                          & \cellcolor[HTML]{8DB4E2}5.93  & \cellcolor[HTML]{8DB4E2}6.71  \\
11 & \cellcolor[HTML]{B1A0C7}1.23 & \cellcolor[HTML]{B1A0C7}1.46 & \cellcolor[HTML]{B1A0C7}1.48 & 3.09                         & 2.81                          & 2.70                          & 4.37                          & 2.53                         & 4.85                          & 3.74                          & \cellcolor[HTML]{8DB4E2}5.23  & \cellcolor[HTML]{8DB4E2}5.11   & 3.30                          & \cellcolor[HTML]{8DB4E2}6.78  & \cellcolor[HTML]{8DB4E2}8.13  \\
12 & \cellcolor[HTML]{B1A0C7}1.32 & 2.06                         & 2.84                         & 3.05                         & 3.57                          & 3.91                          & 3.90                          & 2.04                         & \cellcolor[HTML]{8DB4E2}6.03  & \cellcolor[HTML]{8DB4E2}5.99  & \cellcolor[HTML]{8DB4E2}6.55  & 3.02                           & 4.76                          & \cellcolor[HTML]{8DB4E2}6.95  & \cellcolor[HTML]{538DD5}32.33 \\
13 & \cellcolor[HTML]{CCC0DA}1.79 & \cellcolor[HTML]{CCC0DA}1.97 & 3.63                         & 4.48                         & 4.01                          & \cellcolor[HTML]{8DB4E2}5.92  & \cellcolor[HTML]{8DB4E2}6.73  & 2.72                         & 3.79                          & \cellcolor[HTML]{538DD5}27.38 & \cellcolor[HTML]{8DB4E2}6.28  & \cellcolor[HTML]{8DB4E2}9.22   & \cellcolor[HTML]{8DB4E2}12.57 & \cellcolor[HTML]{538DD5}72.23 & \cellcolor[HTML]{538DD5}27.10 \\
14 & 2.72                         & 2.24                         & \cellcolor[HTML]{8DB4E2}5.94 & 4.43                         & \cellcolor[HTML]{538DD5}24.81 & \cellcolor[HTML]{8DB4E2}8.52  & \cellcolor[HTML]{538DD5}25.18 & 4.85                         & \cellcolor[HTML]{8DB4E2}5.30  & \cellcolor[HTML]{538DD5}55.04 & \cellcolor[HTML]{538DD5}58.81 & \cellcolor[HTML]{538DD5}109.59 & \cellcolor[HTML]{538DD5}43.82 & \cellcolor[HTML]{538DD5}43.84 & \cellcolor[HTML]{538DD5}48.41 \\
15 & 2.96                         & \cellcolor[HTML]{CCC0DA}1.85 & \cellcolor[HTML]{8DB4E2}6.42 & 2.43                         & \cellcolor[HTML]{8DB4E2}8.67  & \cellcolor[HTML]{8DB4E2}14.61 & \cellcolor[HTML]{538DD5}37.10 & 4.97                         & \cellcolor[HTML]{8DB4E2}10.09 & \cellcolor[HTML]{8DB4E2}7.39  & \cellcolor[HTML]{538DD5}24.11 & \cellcolor[HTML]{538DD5}22.31  & \cellcolor[HTML]{538DD5}24.19 & \cellcolor[HTML]{538DD5}24.04 & \cellcolor[HTML]{538DD5}36.50
	\end{tabular}
 }
\end{table}

\section{Conclusion}\label{sec:con}

%This paper focuses on the problem of computing images for quantum transition systems. We present efficient algorithms for identifying a basis for a given subspace and propose various partition schemes for both subspaces and quantum circuits. Additionally, we introduce algorithms for estimating a superset of a subspace, represented as a set of tensor product basis. Finally, we combine these components to form several algorithms capable of performing image computation tasks efficiently for quantum systems.

%In future work, we will consider using our method to do the model checking of quantum systems, especially checking the properties that can be described by temporal logic. We will also consider the image computation of more sophisticated systems, especially systems that can not be represented as quantum circuits.

Image computation plays a key role in model checking classical and quantum transition systems. This paper focuses on advancing the development of model checking for quantum systems by introducing efficient image computation algorithms. We represent quantum circuits as tensor networks and leverage the properties of tensor networks and tensor decision diagrams to design efficient image computation algorithms. Empirical evaluation demonstrates that the contraction partition-based algorithm can greatly improve the efficiency of image computation for quantum transition systems.

% \section*{Acknowledgment}

% \section*{References}

\bibliographystyle{IEEEtran}

\bibliography{reference}

\end{document}